\documentclass[prb,aps]{revtex4}

\usepackage{amsmath}
\usepackage{amssymb}
\usepackage{graphicx}
\usepackage{epsfig}
\usepackage{amsfonts}
\usepackage{color}

\begin{document}
\title{Incommensurate phases of a bosonic two-leg ladder under a flux}
\author{E. Orignac}
\affiliation{{Univ Lyon, Ens de Lyon, Univ Claude Bernard, CNRS, Laboratoire de Physique, F-69342 Lyon, France}}
\author{R. Citro}
\affiliation{Dipartimento di Fisica "E.R. Caianiello", Universit\`a
  degli Studi di Salerno, Via Giovanni Paolo II 132, I-84084 Fisciano
  (Sa), Italy}
\affiliation{Spin-CNR, Via Giovanni Paolo II, 84084 Fisciano, Italy} 
\author{M. Di Dio}
\affiliation{CNR-IOM-Democritos National Simulation Centre, UDS Via Bonomea 265, I-34136, Trieste, Italy}
\author{S. De Palo}
\affiliation{CNR-IOM-Democritos National Simulation Centre, UDS Via Bonomea 265, I-34136, Trieste, Italy}
\affiliation{Dipartimento di Fisica Teorica, Universit\`a Trieste,
  Trieste, Italy}
\author{M.-L. Chiofalo}
\affiliation{Dept. of Physics "Enrico Fermi" and INFN,  Universit\`a di
  Pisa Largo Bruno Pontecorvo 3 I-56127 Pisa, Italy}

\begin{abstract}
A boson two--leg ladder in the presence of a synthetic magnetic flux is investigated by means of bosonization
techniques and Density Matrix Renormalization Group (DMRG).  We follow
the quantum phase transition from the commensurate
Meissner to the incommensurate vortex phase with increasing flux at different fillings.
When the applied flux is $\rho \pi$ and close to it, where $\rho$ is the filling per rung, we find
 a second incommensuration in the vortex state that affects physical
 observables such as the momentum distribution, the rung-rung
 correlation function and the  spin-spin and charge-charge static structure factors.
\end{abstract}
\date{May 23, 2016}
\maketitle

A remarkable characteristic of charged systems with broken $U(1)$
global gauge symmetry such as superconductors is the Meissner-Ochsenfeld
effect\cite{tinkham_book_superconductors}.  In the Meissner phase, below
the critical field $H_{c1}$, a superconductor behaves as a perfect
{diamagnet}, i.e. it develops surface currents that fully screen the
external magnetic field. In a type-II superconductor, for fields above
$H>H_{c1}$, an Abrikosov vortex lattice phase is formed in the system, where
{  the magnetic field penetrates into vortex cores}. In
quasi one-dimensional systems, analogues of the Meissner and Abrikosov
vortex lattice have been predicted for the bosonic two-leg
ladder\cite{kardar_josephson_ladder,orignac01_meissner,cha2011,tokuno2014},
the simplest system where orbital magnetic field effects are allowed.
It was shown that in this model, the quantum phase transition between the Meissner
and the Vortex phase is a commensurate-incommensurate transition\cite{japaridze_cic_transition,pokrovsky_talapov_prl,schulz_cic2d}.
For ladder systems at commensurate filling, a chiral Mott insulator
phase with currents circulating in loops commensurate with the ladder
was obtained\cite{dhar2012,dhar2013,petrescu2013,petrescu2015}
Initially, Josephson junction arrays\cite{vanoudenaarden_josephson_mott,vanoudenaarden_josephson_localization,fazio_josephson_junction_review,lehur2015} were proposed
as experimental realizations of bosonic one dimensional
systems.\cite{bradley_josephson_chain,glazman_josephson_1d}
However, Josephson junctions are dissipative and open
systems\cite{korshunov_dissipative_josephson1d,korshunov_dissipative_josephson1d,bobbert_dissipative_josephson1d_a,bobbert_dissipative_josephson1d_b}
that cannot be described using a  {Hermitian many-body
Hamiltonian in a canonical formalism.}
Moreover, the quantum effects in
the vortex phase of the Josephson ladder are weak\cite{bruder_josephson1d}.
Fortunately, with the recent advent of ultracold atomic gases, another route to realize low dimensional strongly interacting bosonic systems has opened\cite{jaksch05_coldatoms,lewenstein07_coldatoms_review,bloch08_manybody}.
Atoms being neutral, it is necessary to find a way to realize
an artificial magnetic flux acting on the ladder.
{Alternatively, one can consider the mapping of
the two-leg ladder bosonic model to a two-component spinor boson model in which the bosons in the upper leg become spin-up bosons and the bosons in the lower leg spin-down bosons. Under such mapping, the magnetic flux of the ladder becomes a spin-orbit coupling for the spinor bosons.
Theoretical proposals to realize either artificial gauge fields or artificial spin orbit coupling have been put
forward\cite{osterloh05_gauge,ruseckas05_gauge}, and an
artificial spin-orbit coupling has been achieved in a cold atoms
experiment\cite{lin2011_soc,galitski2013_soc}}
Recently, the Meissner effect and the formation of a vortex state
have been observed for non-interacting ultracold bosonic atoms bosons
on a two leg ladder in
artificial gauge fields induced by laser-assisted
tunneling\cite{atala2014}. The behavior of the chiral current as a
function of the coupling strength along the rungs of the ladder,
indicates a diamagnetic phase when it reaches a saturated maximum and
a vortex lattice phase when it starts to decrease.
This experimental achievement has revived the theoretical interest for bosonic ladders
in the presence of magnetic flux and its spinor-boson equivalent in the presence of interaction, where an even richer phase diagram is expected\cite{tokuno2014,zhao2014,keles2015,xu2014,piraud2014,barbiero2014,peotta2014,sterdyniak2014,greschner2015a}.

In the present manuscript, we study the commensurate-incommensurate transitions of the
hard-core boson ladder with equal
densities in the two legs, for varying
interleg coupling and flux\cite{our_2015} and fixed fillings away from
half-filling. {We confirm that above a threshold in the interleg
coupling, the Meissner phase is stable for all fluxes\cite{piraud2014b} while below that threshold the
commensurate-incommensurate phase transition\cite{orignac01_meissner} to
the vortex phase takes place at large enough flux. However, within the vortex phase, we find that a second
incommensuration\cite{our_2015} appears at a flux
commensurate with the filling, which we characterize by different observables.  
}
The paper is organized as follows. In  Sec.~\ref{sec:model}, we present
the model and the Hamiltonian and define the observables. In
Sec.~\ref{sec:boso} we describe the bosonization treatment,
the Meissner state and the commensurate-incommensurate (C-IC) transition.  In
Sec.~\ref{sec:lambda-pi}, we discuss the second
incommensuration as a function of the filling. Finally, in the
conclusion we present the phase-diagram emerging for the half-filled
case.

\section{Model and Hamiltonian}
\label{sec:model}

\begin{figure}
\centering\includegraphics[width=0.7\linewidth]{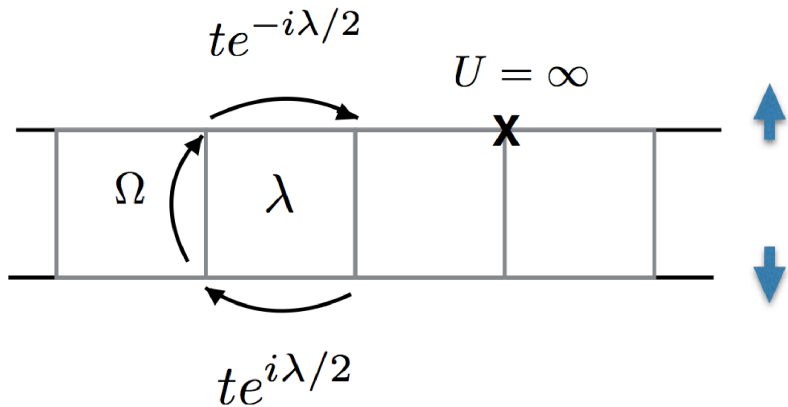}
\caption{Schematic representation of the Hamiltonian Eq.~\ref{eq:full-lattice-ham}.
The presence of an artificial magnetic flux $\lambda$ per plaquette, induces the
hopping terms on the chain to acquire a phase that depends on the spin (chain).
No double occupancy is allowed due to hard-core interaction.}
\label{fig:model}
\end{figure}

The lattice Hamiltonian of the bosonic ladder in a flux\cite{kardar_josephson_ladder,
orignac01_meissner} reads:
\begin{eqnarray}
  \label{eq:full-lattice-ham}
  H_\lambda = \sum_{j,\sigma} -t (b^\dagger_{j,\sigma} e^{i \lambda \sigma}
  b_{j+1,\sigma} +b^\dagger_{j+1,\sigma}  e^{-i \lambda  \sigma}
  b_{j,\sigma})  - \Omega \sum_{j,\sigma} b^\dagger_{j,\sigma} b_{j,-\sigma},
\end{eqnarray}
where the operator
$b^{(\dagger)}_{j,\sigma}$ destroys (creates) a hard core boson on site
$j$ of the $\sigma$ chain. We have defined $\sigma=\pm 1/2$ as the chain
index\cite{orignac01_meissner,our_2015}, $\lambda$ as the flux in each
plaquette (corresponding to a Landau gauge with the vector potential
parallel to the legs), $\Omega$ as the interchain hopping.
The $t e^{i \lambda \sigma}$ is the hopping amplitude on the chain $\sigma$. A schematic picture of the model and its relevant parameters is shown in Fig.\ref{fig:model}.
This hard-core boson model can be mapped into a spin-ladder model with Dzialoshinskii-Moriya
interactions\cite{dzyalo_interaction,moryia_asym_int}, as detailed in Appendix
\ref{sec:spinladder}.
As a result of translational invariance and parity, the spectrum of the
Hamiltonian~(\ref{eq:full-lattice-ham}) is even and $2\pi$-periodic in $\lambda$.

The leg-current operator
$J_\parallel(j,\lambda)$ is defined as:
\begin{eqnarray}
  \label{eq:spin-current}
  J_\parallel(j,\lambda) =  \sum_{\sigma} -it \sigma (b^\dagger_{j,\sigma} e^{i \lambda \sigma}
  b_{j+1,\sigma} - b^\dagger_{j+1,\sigma}  e^{-i \lambda  \sigma}
  b_{j,\sigma}) = \frac{\partial H_\lambda}{\partial \lambda},
\end{eqnarray}
while the rung current is defined as:
\begin{eqnarray}
  \label{eq:spin-y-current}
  J_{\perp}(j)=-i \Omega (b^\dagger_{j,\uparrow} b_{j_\downarrow} -
  b^\dagger_{j,\downarrow} b_{j_\uparrow}).
\end{eqnarray}

The average densities of bosons are $\rho_\sigma=\frac{N_\sigma}
L $ where $N_\sigma$ is the number of particles in chain  $\sigma $  and $L$ is the length of the
chain. In the rest of the manuscript, we will be considering a fixed total
density $\rho= \rho_\uparrow + \rho_\downarrow$. In the absence of applied flux
$\lambda$ the ground state of the system is a rung-Mott Insulator for $\rho=1$ and
a superfluid for $\rho<1$ \cite{crepin2011}. This situation is not changed at finite
$\lambda$ so that for $\rho=1$ Mott-Meissner and Mott-Vortex
phase\cite{petrescu2013,petrescu2015,piraud2014b} are obtained.

For our analysis, we are interested in the
following observables: the rung-current correlator $C(k)$
\begin{eqnarray}
  \label{eq:rung-corr-def}
  C(k)=\sum_j \langle  J_{\perp}(j)  J_{\perp}(0)\rangle e^{-ikj},
\end{eqnarray}
the leg-symmetric density correlator $S_c(k)$
\begin{eqnarray}
  \label{eq:sym-density-corr-def}
  S_c(k)= \sum_{j,\sigma,\sigma'} \langle  n_{j,\sigma}  n_{0,\sigma'}
  \rangle e^{-ikj},
\end{eqnarray}
the leg-antisymmetric density correlator $S_s(k)$
\begin{eqnarray}
  \label{eq:asym-density-corr-def}
  S_s(k)= \sum_{j,\sigma,\sigma'} \sigma \sigma' \langle  n_{j,\sigma}  n_{0,\sigma'}
  \rangle e^{-ikj},
\end{eqnarray}
and the leg-resolved momentum distribution $n_\sigma(k)$
\begin{eqnarray}
  \label{eq:mom-dist-def}
  n_\sigma(k) = \sum_j \langle b_{j,\sigma}^\dagger b_{0,\sigma}
  \rangle e^{-ikj}.
\end{eqnarray}
The non leg-resolved momentum distribution is
$n(k)=n_\uparrow(k)+n_\downarrow(k)$. The latter quantity is
accessible in time-of-flight spectroscopy\cite{atala2014}.

\section{Bosonization of the two-leg boson ladder}
\label{sec:boso}

We apply Haldane's bosonization of interacting bosons\cite{haldane_bosons} to the Hamiltonian~(\ref{eq:full-lattice-ham}) assuming that $\Omega$ is a perturbation.
In the absence of interchain couplings and spin-orbit coupling, the Hamiltonian of the bosons can be written as:

\begin{eqnarray}
  \label{eq:2cll-ham}
  H_0=\sum_\sigma \int \frac{dx}{2\pi} \left[ u_\sigma K_\sigma (\pi  \Pi_\sigma)^2
     + \frac{u_\sigma}{K_\sigma} (\partial_x\phi_\sigma)^2  \right],
\end{eqnarray}
  where $[\phi_\alpha(x),\Pi_\beta(x')]=i\delta_{\alpha\beta}
\delta(x-x')$, $u_\sigma$ is the velocity of excitations, $K_\sigma$
is the Tomonaga-Luttinger (TL) exponent. In the case of hard-core
bosons, $u_\sigma = 2t \sin (\pi \rho^{0}_\sigma)$ and $K_\sigma=1$.

Introducing the fields $\theta_\alpha =\pi \int^x
\Pi_\alpha$, we can represent\cite{haldane_bosons} the boson annihilation operators as:
\begin{eqnarray}
  \label{eq:boson-annihil}
  \frac{b_{j\sigma}}{\sqrt{a}}=\psi_\sigma(x)=e^{i\theta_\sigma(x)} \sum_{m=0}^{+\infty} A^{(\sigma)}_m \cos
  (2m \phi_\sigma(x) -2m \pi \rho^{(0)}_\sigma x),
\end{eqnarray}
and the density operators\cite{haldane_bosons} as:
\begin{eqnarray}
  \label{eq:boson-density}
  \frac{n_{j\sigma}}{a}=\rho_\sigma(x)=\rho_\sigma^{(0)} -\frac 1 {\pi} \partial_x
  \phi_\sigma + \sum_{m=1}^\infty B_m^{(\sigma)} \cos  (2m
  \phi_\sigma(x) -2m \pi \rho^{(0)}_\sigma x).
\end{eqnarray}
Here, we have introduced the lattice spacing $a$, while $A_m$ and $B_m$
are non-universal coefficients. In the case of hard core bosons at
half filling, these coefficients have been found
analytically\cite{ovchinnikov2004}.
From Eq.(\ref{eq:boson-annihil}), we deduce the bosonized expression
of the interchain hopping as:
\begin{eqnarray}
  \label{eq:boso-hopping}
  H_{\text{hop.}} = - \Omega A_0^2 \int dx \cos (\theta_\uparrow -\theta_\downarrow),
\end{eqnarray}
where we have kept only the most relevant term in the renormalization
group sense\cite{orignac01_meissner}.

For a model with equivalent up and down leg as
Eq. (\ref{eq:full-lattice-ham}), and in the absence of the spontaneous
density imbalance between the chains found for weak
repulsion\cite{uchino2015,uchino2016}, $u_\uparrow=u_\downarrow$ and
$K_\uparrow=K_\downarrow$, it is 
convenient to introduce the leg-symmetric and leg-antisymmetric representation:

\begin{eqnarray}
  \label{eq:rotation}
  \Pi_c=\frac 1 {\sqrt{2}} (\Pi_\uparrow + \Pi_\downarrow) &&
  \Pi_s=\frac 1 {\sqrt{2}} (\Pi_\uparrow - \Pi_\downarrow) \\
 \phi_c=\frac 1 {\sqrt{2}} (\phi_\uparrow + \phi_\downarrow) &&
\phi_s=\frac 1 {\sqrt{2}} (\phi_\uparrow - \phi_\downarrow),
\end{eqnarray}

in order to rewrite (\ref{eq:2cll-ham})--~(\ref{eq:boso-hopping}) as:
\begin{eqnarray}
  \label{eq:spin-charge}
  H&=&H_c+H_s \\
\label{eq:charge-luttinger}
H_c&=&\int \frac{dx}{2\pi} \left[ u_c K_c (\pi \Pi_c)^2 +
  \frac{u_c}{K_c} (\partial_x \phi_c)^2\right] \\
\label{eq:spin-sine-gordon}
H_s&=&\int \frac{dx}{2\pi} \left[ u_s K_s (\pi \Pi_s)^2 +
  \frac{u_s}{K_s} (\partial_x \phi_s)^2\right] - \Omega A_0^2 \int  dx
\cos (\sqrt{2}\theta_s).
\end{eqnarray}
{The Hamiltonian $H_c$ describes the gapless leg-symmetric density
modes, while $H_s$, which describes the leg-antisymmetric modes, has
the form of a quantum sine-Gordon
model\cite{coleman_equivalence,luther_solitons,rajaraman_instanton}
and is gapful for $K_s>1/4$. In a model of bosons with spin-orbit
coupling, $H_s$ would describe the spin modes, and $H_c$ the total
density (\textit{i.e.} the ``charge'' in the bosonization litterature)
modes. }
Till now, we haven't considered the effect of the flux $\lambda$. We
now show that it can be  exactly incorporated in the bosonized Hamiltonian.
In the absence of interchain hopping $\Omega$, we
can perform independent gauge transformations on the upper and the
lower leg of the ladder. In particular, the gauge transformation:
\begin{eqnarray}
  \label{eq:gauge}
  b_{j,\sigma} = e^{-i \lambda \sigma j} \bar{b}_{j,\sigma}
\end{eqnarray}
entirely removes $\lambda$ from the Hamiltonian. We can then apply the
Haldane bosonization~(\ref{eq:2cll-ham})--(\ref{eq:boson-annihil}) to the
$\bar{b}_{j,\sigma}$ operators. Combining the resulting expressions
with Eq.~(\ref{eq:gauge}), we see that $b_{j,\sigma}$ has now a bosonized
expression of the form:
\begin{eqnarray}
  \frac{b_{j\sigma}}{\sqrt{a}}=\psi_\sigma(x)=e^{i\bar{\theta}_\sigma(x)-i \sigma \frac{\lambda x}{a} } \sum_{m=0}^{+\infty} A^{(\sigma)}_m \cos
  (2m \bar{\phi}_\sigma(x) -2m \pi \rho^{(0)}_\sigma x).
\end{eqnarray}

The boson operators $b_{j,\sigma}$ can be written
in the form (\ref{eq:boson-annihil}) with $\phi_\sigma =
\bar{\phi}_\sigma$ and $\theta_\sigma(x)=\bar{\theta}_\sigma(x)-
\sigma \frac{\lambda x}{2 a}$ and the Hamiltonian expressed in terms of
the fields $\Pi$ and $\phi_\sigma$ now reads:


 \begin{eqnarray}
    H=\sum_\sigma \int \frac{dx}{2\pi} \left[ u_\sigma K_\sigma \left(\pi
      \Pi_\sigma + \sigma \frac{\lambda}{a}\right)^2
     + \frac{u_\sigma}{K_\sigma} (\partial_x \phi_\sigma)^2  \right], 
\end{eqnarray}
leading to a modified Hamiltonian for the leg antisymmetric modes, 
\begin{eqnarray}\label{eq:flux-sg-ham}
  H_s=\int \frac{dx}{2\pi} \left[ u_s K_s \left(\pi \Pi_s +\sqrt{2}
      \frac \lambda a \right)^2 +
  \frac{u_s}{K_s} (\partial_x \phi_s)^2\right] - \Omega A_0^2 \int  dx
\cos (\sqrt{2}\theta_s).
\end{eqnarray}
As discussed in Ref.~\onlinecite{orignac01_meissner}, when $\Omega \ne 0$ the
$\lambda$ term is imposing a gradient of $\theta_\uparrow
-\theta_\downarrow$, while the term~(\ref{eq:boso-hopping}) is imposing a
constant value of $\theta_\uparrow -\theta_\downarrow$.  For sufficiently large values of $\lambda$ it becomes
energetically advantageous to populate the ground state with solitons
giving rise to an incommensurate phase. In the ladder language, such
incommensurate phase is the vortex lattice\cite{orignac01_meissner}.

\subsection{Gapful excitations in the Meissner state}
\label{sec:sol-breathers}

The
quantum sine-Gordon model~(\ref{eq:spin-sine-gordon}) is
integrable\cite{zamolodchikov79_smatrices,dorey_smatrix_review} and
its spectrum is fully determined.
The Hamiltonian~(\ref{eq:spin-sine-gordon}) for $\lambda=0$  has a  gap $\Delta_s \sim
\frac {u_s}{a} |a \Omega/u_s|^{\frac{2K_s}{4K_s-1}}$, {where
$a$ is the lattice spacing},  for $K_s>1/4$. In
its ground state $\langle \theta_s\rangle \equiv 0 [\pi \sqrt{2}]$.
For $1/4<K_s<1/2$, the excitations above the ground state are solitons
and antisolitons with the relativistic dispersion
$E_s(k)=\sqrt{(u_sk)^2 + \Delta_s^2}$. The soliton and the antisoliton
are topological excitations of the field $\theta_s$ that carry a leg
current $j_s^z=\pm u_s K_s$.  In the case where one is considering
the gap between the ground state and an excited  state of total spin
current zero (i. e. containing at least one soliton and one
antisoliton), the measured gap will be $2\Delta_s$.
When $K_s>1/2$, the solitons\cite{luther_solitons}
and the antisolitons attract each other and can form bound states
called breathers that do not carry any spin current. The measured gap
between the ground state and the lowest zero current state will be the
mass of the lightest breather\cite{zamolodchikov_energy_sg}
\begin{eqnarray}
  \label{eq:1st-breather-mass}
  \Delta_s = \frac{4 u_s}{a \sqrt{\pi}} \frac{\Gamma\left(\frac 1
      {8K_s-2}\right) \sin\left(\frac \pi
      {8K_s-2}\right) }{\Gamma\left(\frac {2K_s}
      {4K_s-1}\right)} \left[\frac{\pi \Gamma\left(1-\frac 1
      {4K_s}\right)}{\Gamma\left(\frac 1
      {4K_s}\right)} \frac{\Omega A_0^2}{2u}
\right]^{\frac{2K_s}{4K_s-1}}.
\end{eqnarray}
In the case of hard core bosons\cite{cazalilla2011}, which is the one considered in the numerical analysis here, we have
$K_c=K_s=1$, so  $\Delta_s \sim \Omega^{2/3}$. In that limit,
the Hamiltonian~(\ref{eq:spin-sine-gordon})
has been studied
in relation with spin-1/2 chain materials with staggered
Dzialoshinskii-Moriya in a magnetic
field\cite{oshikawa_cu_benzoate_short,oshikawa_cu_benzoate,essler99_cu_benzoate,essler03_cubenzoate,nojiri_breather,umegaki2012}.
With a weak spin-spin repulsion  logarithmic
corrections\cite{oshikawa_cu_benzoate_short,oshikawa_cu_benzoate}
are actually obtained as a result a marginal flow, and
$\Delta_s \sim  \Omega^{2/3}|\ln\Omega|^{1/6}$.
Besides the solitons and antisolitons, there are two
breathers\cite{uhrig_excitation_staggered,affleck86_dimerized,tsvelik_excitation_staggered},
a light breather of mass $\Delta_s$ and a heavy breather of mass
$\sqrt{3} \Delta_s$.

The amplitude $A_0$ in Eq.~(\ref{eq:1st-breather-mass}) can be estimated for hard core bosons in the case
of low density, using the continuum limit\cite{vaidya_bose1d_exact,Gangardt_correlations} or in the case of
half-filling\cite{ovchinnikov2004}.
In the first case,$\frac{A_0^2 a} 2 = \frac{G(3/2)^4 n_0^{1/2}}{\sqrt{2\pi}}$
where $G$ is the Barnes G function and $n_0$ is the number of
particles per site, while in the second case, $\frac{A_0^2 a} 2 \simeq
0.588352$.
This gives the estimates:
\begin{eqnarray}
  \label{eq:gap-low-density}
  \Delta_s=\frac{u_s}{a} \frac{2 \Gamma(1/6)}{\sqrt{\pi}
    \Gamma(2/3)} \left(\frac{\sqrt{\pi} \Gamma(3/4)G(3/2)^4}{\sqrt{2} \Gamma(1/4)}
    \frac{\Omega  a}{u_s}\right)^{2/3} n_0^{1/3},
\end{eqnarray}
for low density, and:
 \begin{eqnarray}
  \label{eq:gap-half-filled}
  \Delta_s=\frac{u_s}{a} \frac{2 \Gamma(1/6)}{\sqrt{\pi}
    \Gamma(2/3)} \left(\frac{2 \sqrt{\pi} \Gamma(3/4) C_0}{ \Gamma(1/4)}
    \frac{\Omega  a}{u_s}\right)^{2/3},
\end{eqnarray}
for half-filling.
\subsection{Correlation functions in the Meissner state}
\label{sec:meissner-state}

As for $K_s>1/4$ the ground state of $H_s$ has $\theta_s$ long-range
ordered and the excitations above the ground state are gapped, the
system described by (\ref{eq:spin-charge}) is a Luther-Emery
liquid\cite{luther_exact}. In such a phase,
\begin{eqnarray}
  \label{eq:spin-gap-annihil}
  b_{j,\sigma} \sim \langle e^{i\sigma \sqrt{2} \theta_s(ja)} \rangle  e^{i\frac{\theta_c(ja)}{\sqrt{2}}} ,
\end{eqnarray}
giving rise to correlations $\langle b_{j, \sigma}
b_{j,\sigma'}\rangle \sim \frac{|\langle \cos  (\theta_s(ja)/\sqrt{2})\rangle|^2}
{|j-j'|^{1/(4K_c)}} $. This behavior is a remnant
of the single condensate obtained in the non-interacting case\cite{atala2014,tokuno2014}.
Since
\begin{eqnarray}
  J_{\perp} = \Omega A_0^2 \sin \sqrt{2} \theta_s,
\end{eqnarray}
we have $\langle J_{\perp}(x) \rangle=0$ and
\begin{eqnarray}
  \langle J_{\perp}(x) J_{\perp}(x') \rangle \sim e^{-|x-x'|/\xi},
\label{rungrung_rs}
\end{eqnarray}
as $|x-x'| \to \infty$. Thus, the average rung-current vanishes and its fluctuations are short ranged and commensurate, so that $C(k)$
takes a Lorentzian shape in the vicinity of $k=0$.

In the case of density-density and spin-spin correlation functions, we
have:
\begin{eqnarray}
  \frac 1 a \sum_\sigma n_{j,\sigma} \sim \rho^{(0)}
-\frac {\sqrt{2}} \pi \partial_x \phi_c + \sum_m B_m \cos (m \sqrt{2} \phi_c -\pi m
    \rho^{(0)} x) \cos (m \sqrt{2} \phi_s),
\end{eqnarray}
and
\begin{eqnarray}
  \frac 1 {2a} \sum_\sigma \sigma n_{j,\sigma} \sim \rho^{(0)}  -\frac
  1 {\pi \sqrt{2}}  \partial_x \phi_s + \sum_m B_m \cos (m \sqrt{2} \phi_c -\pi m
    \rho^{(0)} x) \sin (m \sqrt{2} \phi_s).
\end{eqnarray}
Since the field $\theta_s$ is long-range ordered,
exponentials $e^{i \beta \phi_s}$ and derivatives $\partial_x^n
\phi_s$ of its dual field $\phi_s$  are short-range
ordered. As a result, the density correlations decay as $(x-x')^{-2}$
at long distance leading to $S_c(k)=K_c |k|/(2\pi)$, while the spin-spin
correlations are decaying exponentially giving a Lorentzian shape for
$S_s(k)$. Finally, if we consider the longitudinal spin current,
the obtained bosonized expression is:
\begin{eqnarray}
  J_\parallel(\lambda)&=& \frac {u_s K_s}{\sqrt{2}} \left( \Pi_s +
     \frac{\lambda}{\pi a \sqrt{2}}\right).
\end{eqnarray}
In the Meissner phase, {the linear behavior is obtained, with} $\langle J_\parallel(\lambda) \rangle = \lambda {u_s
  K_s }({2\pi a})^{-1}$.

\subsection{ Commensurate-Incommensurate transition}
\label{sec:vortex}

Adding the spin-orbit coupling $\lambda$ in
(\ref{eq:spin-sine-gordon}) gives a Hamiltonian for the
spin modes:
\begin{eqnarray}\label{eq:spin-orbit-sinegordon}
  H_s=\int \frac{dx}{2\pi} \left[ u_s K_s \left(\pi \Pi_s + \frac{\lambda} {\sqrt{2} a} \right)^2 + \frac {u_s}{K_s} (\partial_x \phi_s)^2\right]
  - \Omega A_0^2 \int dx \cos \sqrt{2} \theta_s.
\end{eqnarray}
Expanding $(\pi \Pi_s + \lambda/{\sqrt{2} a})^2$ and using $\pi
\Pi_s=\partial_x \theta_s$, up to a constant shift, the spin orbit coupling adds a term:
\begin{eqnarray}
  +\frac{u_s K_s \lambda } {a\sqrt{2}}  \int
  \frac{dx}{\pi} \partial_x \theta_s,
\end{eqnarray}
to the Hamiltonian~(\ref{eq:spin-sine-gordon}).

Now, if we  call $N_s$ is the number of sine-Gordon
solitons and   $N_{\bar{s}}$ the number of antisolitons, we have:
\begin{eqnarray}
  \label{eq:soliton-count}
  N_s-N_{\bar{s}} = \int_{-\infty}^\infty \frac{dx}{\pi
    \sqrt{2}} \partial_x \theta_s,
\end{eqnarray}
and the contribution of the spin-orbit coupling is rewritten
\begin{eqnarray}\label{eq:soliton-chemical-pot}
  \frac{u_s K_s \lambda}{a} ( N_s-N_{\bar{s}}),
\end{eqnarray}
showing that $\lambda$ acts as a chemical potential for solitons or
antisolitons. On the other hand, the energy cost of forming $N_s$ solitons and $N_{\bar{s}}$
antisolitons is $\Delta_s (N_s+N_{\bar{s}})$.
When $|\lambda|>\lambda_c=\frac{a \Delta_s}{u_s K_s}$, there is an
energy gain to create solitons (or antisolitons depending on the sign of $\lambda$)  in the ground
state. {Because of the fermionic character of
solitons\cite{haldane1982}, their density remains
finite, and we obtain another Luttinger liquid. This is the
commensurate-incommensurate
transition\cite{japaridze_cic_transition,pokrovsky_talapov_prl,schulz_cic2d,papa01_exponents}.
A detailed picture can be obtained for $K_s=1/2$, where solitons can
be treated as non-interacting fermions as discussed in
App.\ref{app:fermionization}. }

In the incommensurate (IC) phase, the Hamiltonian describing the Luttinger
liquid of solitons is:
\begin{eqnarray}
  \label{eq:soliton_ll}
  H=\int \frac{dx}{2\pi} \left[ u_s^*(\lambda) K_s^*(\lambda) (\pi
    \hat{\Pi_s})^2 + \frac  {u_s^*(\lambda)} {K_s^*(\lambda)} (\partial_x
    \phi_s)^2\right],
\end{eqnarray}
with $\theta_s=\hat{\theta}_s -\mathrm{sign}(\lambda) q(\lambda)
x/\sqrt{2}$. The density of solitons is proportional to $q(\lambda)$,
while $u_s^*(\lambda)$ is the renormalized velocity of excitations and
$K_s^*(\lambda)$ is the renormalized Luttinger exponent.

We now address the behavior of the observables in the IC phase.
Near the transition\cite{schulz_cic2d,chitra_spinchains_field}, for
$\lambda \to \lambda_c+0$, $K_s^*(\lambda) \to 1/2$, $ q(\lambda)
\rangle \propto \sqrt{\lambda -\lambda_c}$ and $u_s^*(\lambda) \propto
\sqrt{\lambda -\lambda_c}$.
The expression of the spin current in the IC phase now becomes:
\begin{equation}
  \label{eq:spin-current-ic-phase}
  \langle J_\parallel(\lambda)\rangle =\frac{ u_s K_s}{2}\left(
    \frac{\lambda}{\pi a} - \mathrm{sign}(\lambda) q(\lambda) \right),
\end{equation}
namely the existence of a finite soliton density reduces the
average spin current. This justifies the interpretation of these solitons as
vortices letting the current to flow along the legs.
For large $\lambda$, we
have $q(\lambda) \sim |\lambda| /(\pi a)$, so that the
expectation value of the spin current eventually vanishes for large flux values.

Let us turn to the momentum distribution. In the IC phase and for finite size $L$ with periodic boundary conditions one has:
\begin{eqnarray}
  \label{eq:corr-ic}
  \langle b^\dagger_{j,\sigma} b_{j,\sigma'}\rangle = \frac{e^
    {i \sigma q(\lambda) (x-x')} \delta_{\sigma \sigma'}
  }{\left[\frac L \pi \sin \left(\frac{\pi |x-x'|} L\right)\right]^{1/(4K_c)+1/(4K_s^*)}}.
\end{eqnarray}
As a result, for $1/(4K_c)+1/(4K_s^*)<1$ one has:
\begin{eqnarray}
  \label{eq:nk-fss}
  n_\sigma(k) = 2 \left(\frac L {2\pi} \right)^{1-\frac 1 {4K_c} -\frac
    1 {4K_s^*}} \frac{\Gamma\left(1-\frac 1 {4K_c} -\frac
    1 {4K_s^*} \right) \cos \left(\frac \pi  {8K_c} +\frac
    \pi {8K_s^*}\right) \Gamma\left(\frac 1 {8K_c} +\frac
    1 {8 K_s^*} + \frac {L |k-\sigma q(\lambda)|}{2\pi}
  \right)}{\Gamma\left(1- \frac 1 {8K_c} - \frac
    1 {8 K_s^*} + \frac {L |k-\sigma q(\lambda)|}{2\pi}  \right)},
\end{eqnarray}
so that now $n_\sigma(k)$ has a peak for $k=\sigma q(\lambda)$, whose height scales as
$L^{1-1/(4K_c)-1/(4K_s^*)}$. That peak becomes a power-law divergence
in the limit of $L\to \infty$. Comparing with the
 non-interacting case\cite{atala2014}, these power-law divergences are
 the remnant of the Bose condensate\cite{tokuno2014} formed at $k=0$
 in the Meissner phase or at $k=\pm q(\lambda)/2$ in the vortex
 phase.

Turning to the spin-current correlation function, in the IC phase  we
have\cite{orignac01_meissner,cha2011}:
\begin{eqnarray}
\langle J_{\perp}(j)
J_{\perp}(j') \rangle \sim \frac{\cos [q (\lambda)
(x-x')]}{\left[\frac L \pi \sin \frac{\pi |j-j'|} L\right]^{\frac 1
  {K_s^*}}}.
\end{eqnarray}
Since $1/2 \le K_s^* \le 1$, the correlation function $C(k)$
presents in the IC phase two cusps at $k=\pm q(\lambda)$.

Turning now to the density correlation function, we have:
\begin{eqnarray}
  \left\langle \sum_{\sigma,\sigma'}  n_{j,\sigma}  n_{0,\sigma'} \right\rangle
  &&
  \sim - \frac{2 K_c}{L^2 \sin^2 \left(\frac {\pi j} L\right)} +
  \frac{\cos (\pi \rho^{(0)} j)}{\left[\frac L \pi \sin \left( \frac{\pi
        j} L \right) \right]^{K_c + K_s^* }}, \\
     \left\langle \sum_{\sigma,\sigma'} \sigma \sigma' n_{j,\sigma}  n_{0,\sigma'} \right\rangle
  &&
  \sim - \frac{K_s}{2 L^2 \sin^2 \left(\frac {\pi j} L\right)} +
  \frac{\cos (\pi \rho^{(0)} j)}{\left[\frac L \pi \sin \left( \frac{\pi
        j} L \right) \right]^{K_c + K_s^* }}.
\end{eqnarray}
Since $1 \le K_c+K_s^* \le 2$, we find,{after taking the Fourier
transform},  that both $S_s(k)$ and
$S_c(k)$ possess  cusp singularities $S_{c/s}(k) \sim S_{c/s}(\pi \rho^{0)}) + C_{c/s} |k-\pi
\rho^{(0)}|^{K_c+K_s^*-1}$ in the vicinity of $k=\pi \rho^{(0)}$ in the vortex
phase, with evident notation for the subscript $c/s$.  In the hard core boson system, with $K_c=K_s^*=1$, the cusp
singularities become slope discontinuities.

Moreover, the behaviors $S_c(k)\sim \frac{2 K_c |k|}{\pi} $ and
$S_s(k) \sim \frac{K_s |k|}{2\pi} $ as
$k\to 0$ signal that both charge and spin excitations are gapless in
the vortex phase.

We performed numerical simulations for the hard--core spinless bosons on a two-leg ladder as a
function of flux and interchain hopping and for different fillings by means of DMRG simulations\cite{white_dmrg,schollwock2005}
with Periodic Boundary Conditions (PBC). Simulations are performed for sizes up to $L=64$, keeping up to $M=1256$ states during the renormalization
procedure. The truncation error, that is the weight of the discarded states, is at most of order $10^{-6}$,
while the error on the ground-state energy is of order $5\times10^{-5}$ at most.

\begin{figure}[h]
\begin{center}
\includegraphics[height=55.5mm]{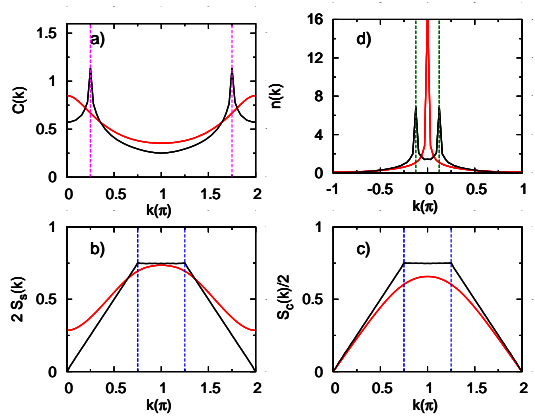}
\end{center}

\caption{First incommensuration: appearance of the standard Vortex phase.
DMRG simulation for $L=64$ in PBC for $\lambda=\pi/4$ and $\rho=0.75$ ($\lambda=\rho\pi$) .
FT of the correlation functions described in the text for two different values of interchain coupling
$\Omega/t=0.0625$ (black solid line) and $0.5$(red solid line), respectively in the Vortex and Meissner phase
Panel a): rung-current correlation function $C(k)$. Panel b): spin correlation function $S_s(k)$ multiplied by a factor $2$. 
Panel c): charge correlation $S_c(k)$ divided by a factor $2$. Panel d): total momentum distribution $n(k)$. See text for the corresponding definitions. The blue dashed lines in panels b) and c) signal the values $k=\pi \rho$ and $k=2\pi-\pi \rho$.
The magenta dashed lines in panel a) signals the peaks positions of $C(k)$, $k=\lambda$ and $k=2\pi-\lambda$. The dark-green dashed lines in panel d) signals the peaks of the momentum distribution at $k=\pm\lambda/2$. }
\label{fig:cic_n_0.75}
\end{figure}

\begin{figure}[h]
\begin{center}
\includegraphics[height=55.5mm]{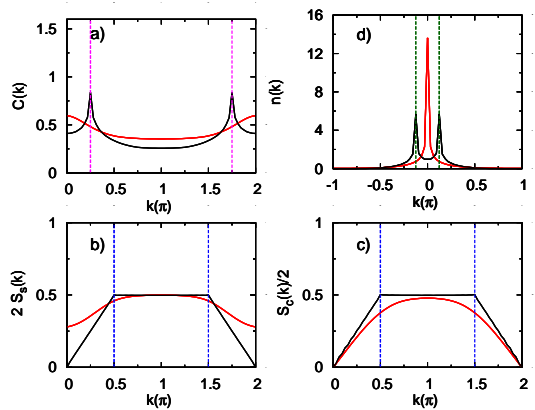}
\end{center}

\caption{The same as in Fig.~\ref{fig:cic_n_0.75} for $\rho=0.5$.}
\label{fig:cic_n_0.5}
\end{figure}

A summary for the behavior of observables and correlation functions
across the commensurate-incommensurate transition at two different
fillings is shown in Fig.~\ref{fig:cic_n_0.75} for $\rho=0.75$ and in
Fig.~\ref{fig:cic_n_0.5} for $\rho=0.5$.  {In both cases, no
spontaneous density imbalance\cite{uchino2015,uchino2016} 
between the chains is present.} In each panel a) of the two
figures we compare the behavior of the Fourier Transform (FT) of the
rung-current correlation function $C(k)$ in the Meissner phase and in the
Vortex phase. The numerical data confirm the prediction of a
structureless shape in the Meissner phase and the appearance of two
cusp--like peaks in the Vortex phase, respectively at $k=q(\lambda)$
and $k = 2\pi-q(\lambda)$.
Since we show data in the vortex phase far from the transition,
$q(\lambda)=\lambda$, as expected. The spin gap closure in the Vortex
phase
is visible also in the low-momentum behavior of the spin static
structure factor $S_s(k)$ displayed in each panel b) of the two
figures~\ref{fig:cic_n_0.75} and~\ref{fig:cic_n_0.5}: in the Vortex phase $S_s(k)=  K_s |k|/2 \pi $ while in the
Meissner phase $S_s(k) = S_s(0)+ a k^2 $ with $S_s(0)>0$.  In these
cases $K_s=1$ as expected for a hard-core boson system.  At large
momenta the Lorentzian profile centered at $k=\pi$, characteristic of the
Meissner phase, is replaced by two slope discontinuities at
$k=\pi \rho$ and $k=2\pi-\pi \rho$ as expected in the Vortex
phase for $K_s=K_c=1$.
The same evolution can be seen in the charge
static structure factors shown in the c) panels of
Figs.\ref{fig:cic_n_0.75} and \ref{fig:cic_n_0.5}.  The
commensurate-incommensurate transition is clearly visible in the
momentum distribution shown in panels d) of
Figs.\ref{fig:cic_n_0.75} and \ref{fig:cic_n_0.5}: in the Meissner phase
it presents only one cusp-like peak at $k=0$ as expected in a
bosonic  Tomonaga-Luttinger liquid, while the Vortex phase it is characterized by
two peaks with same shape, centered at $k=\pm q(\lambda)/2$.

For $K_s=1/2$, the sine-Gordon
Hamiltonian~(\ref{eq:spin-orbit-sinegordon}) can be rewritten as a
free fermion Hamiltonian\cite{coleman_equivalence,luther_exact}
allowing a more detailed treatment of the commensurate-incommensurate
transition\cite{japaridze_cic_transition,orignac01_meissner}. Such a treatment sheds additional light on the physics of this
commensurate-incommensurate transition, providing an overall alternative description considering that no differences are expected at a qualitative level away from the $K_s=1/2$ case. The details of such derivation are accounted for in Appendix~\ref{app:fermionization}.

\section{The second incommensuration appearing at $\lambda \simeq \pi \rho$}
\label{sec:lambda-pi}

As $\lambda$ gets close to $\pi \rho$, with $\rho=N/L$ is the density per
rung and for $N/L$ not small compared to unity, $\lambda u_s/a$ becomes of the order of
the energy cutoff $u_s/a$ and the form
~(\ref{eq:spin-orbit-sinegordon}) for the
Hamiltonian cannot be used. In order to describe the low--energy
physics at $\lambda=\rho \pi$,
it is necessary to choose a gauge with the vector potential
along the rungs of the ladder, so that the interchain hopping reads:
\begin{equation}
  \label{eq:gauge-rung}
  H_{hop.}=\Omega \sum_{j,\sigma} e^{i 2\pi\sigma \left( \frac{N}{L}\right)j} b^\dagger_{j,\sigma} b_{j,-\sigma}.
\end{equation}

Applying bosonization to (\ref{eq:gauge-rung}), we obtain from
(\ref{eq:boson-annihil})  the
following representation for the interchain hopping:
\begin{eqnarray}
  \label{eq:rung-pi}
  H_{hop}=\frac{\Omega}{2\pi a} \int dx e^{i\sqrt{2} \phi_c} \left[
    e^{-i\sqrt{2}(\theta_s +\phi_s)} +e^{-i\sqrt{2}(\theta_s -\phi_s)}\right]
+ e^{-i\sqrt{2}\phi_c} \left[e^{i\sqrt{2}(\theta_s +\phi_s)} +e^{i\sqrt{2}(\theta_s
  -\phi_s)}\right].
\end{eqnarray}
The latter can be rewritten in terms of $\mathrm{SU(2)}_1$
Wess-Zumino-Novikov-Witten (WZNW) currents \cite{gogolin_1dbook}:
\begin{eqnarray}
  \label{eq:rung-pi-wzw}
  H_{hop}=\Omega \int dx \left[ i e^{i\sqrt{2} \phi_c} (J_R^- + J_L^-) +
  \text{H. c} \right].
\end{eqnarray}
In the case of $N/L=1$, the complex exponential of
Eq.(\ref{eq:rung-pi-wzw})  is replaced\cite{our_2015} by a
cosine $\cos \sqrt{2} \phi_c$. At commensurate fillings, $N/L$ is a
rational number $p/m$ with $p,m$ mutually prime and a
term $\cos m \sqrt{2} \phi_c$ is also present in the Hamiltonian. In
the presence of such term, the symmetry of the U(1) charge
Hamiltonian is lowered to $\Bbb{Z}_m$ and a spontaneous symmetry
breaking giving rise to a charge gap and a long-range ordered
$e^{i\sqrt{2} \phi_c}$  becomes possible { in the presence of
long ranged interactions\cite{schulz_mott_revue}. In such
case, an insulating phase with a second incommensuration is obtained.\cite{our_2015}  }

{ At generic filling, or when the term $\cos m \sqrt{2} \phi_c$ is irrelevant , we have
an unbroken $U(1)$ symmetry $\phi_c \to \phi_c+\gamma$ and $\theta_s
\to \theta_s +\gamma$. In such case, the Mermin-Wagner
theorem\cite{mermin_wagner_theorem,hohenberg67_theorem} precludes long
range ordering for $\phi_c +\theta_s$. However, since the perturbation
in (\ref{eq:rung-pi-wzw}) is {relevant in the renormalization group
sense} and has non-zero conformal spin, it is still
expected to give rise to incommensurate correlations {at the strong
  coupling fixed point}.
To give a
qualitative picture of such incommensuration, we turn to a mean-field
treatment. {Compared with the half-filled case, the
assumption $\langle \phi_c\rangle =\gamma$  would correspond
to a spontaneously broken $U(1)$ symmetry, not
permitted by the Mermin-Wagner theorem. The Gaussian
fluctuations of the $\phi_c$ modes around the saddle point would in
fact restore  the $U(1)$ symmetry that one has to assume broken to use a
mean-field theory. To \emph{partially} take into account  the effect
of these
fluctuations, we will first solve the mean-field theory for an
arbitrary value of $\gamma$, and we will then average the
obtained correlation functions over $\gamma$. Such averaging procedure
ensures that $\langle e^{i\sqrt{2} \phi_c}\rangle = 0$, and more
generally that the obtained correlation functions respect the $U(1)$
symmetry of the Hamiltonian.
Of course, that procedure is not expected to produce
quantitative estimates, since the fluctations of $\phi_c$ are
underestimated. In particular, the amplitude of the incommensuration
can be less than the one expected from the mean field theory, and the
decay exponents of the correlations can be larger. But the mean field
treatment is providing some insight on the correlation functions that
can reveal the presence of a second incommensuration at the fixed
point.}
Assuming $\langle \phi_c\rangle =\gamma$, after the transformation
$\theta_s \to \theta_s+\gamma$ and $\phi_c \to \phi_c+\gamma$  the Hamiltonian $H_c+H_s+H_{hop}$
can be treated in mean-field
theory\cite{nersesyan_incom,lecheminant2001,jolicoeur2002,zarea04_chiral_xxz},
After defining :
\begin{eqnarray}
  \label{eq:rung-pi-selfcon}
  \frac{g_c}{\pi a} &=& 8 \Omega \langle J_R^y + J_L^y \rangle_{s,MF},\nonumber \\
  h_s &=&  8 \Omega \langle \cos \sqrt{2} \phi_c\rangle_{c,MF},
\end{eqnarray}
using a $\pi/ 2$ rotation around the $x$ axis, $J_{\nu}^y
=\tilde{J}_{\nu}^z$, $J_\nu^z=- \tilde{J}_{\nu}^y$, and applying
abelian bosonization\cite{nersesyan_2ch}, we rewrite:
\begin{eqnarray}
  \label{eq:rung-pi-mf-rotated}
  H_s^{MF}&=&\int \frac{dx}{2\pi} u_s \left[(\pi \tilde{\Pi}_s)^2 +
    (\partial_x \tilde{\phi}_s)^2 \right] - \frac {h_s}{\pi \sqrt{2}}
  \int \partial_x \tilde{\phi}_s dx ,
\end{eqnarray}
which allows us to write:
\begin{eqnarray}
  - \frac {1}{\pi \sqrt{2}} \langle \partial_x \tilde{\phi}_s\rangle =
  \sum_{\nu=R,L} \langle \tilde{J}_{\nu}^z \rangle =  \langle J_R^y +
  J_L^y  \rangle = -\frac {h_s}{2\pi u_s}.
\end{eqnarray}
In turn, this allows us to solve (\ref{eq:rung-pi-selfcon}) with $h_s \sim
\Omega^2$ and $g_c \sim \Omega^3$. We obtain a gap in the total
density excitations, $\Delta_c \sim \Omega^2$, while the antisymmetric
modes remain gapless and develop an incommensuration. To characterize
the incommensuration, we first make a shift of the field
$\tilde{\phi}_s \to  \tilde{\phi}_s + \frac{h_s x}{u_s \sqrt{2}}$ while $\tilde{\theta}_s \to  \tilde{\theta}_s $, thus
\begin{eqnarray}
  \label{eq:rotation-primaries}
  \sin \sqrt{2} \theta_s &=& \sin \sqrt{2} \tilde{\theta}_s \\
  \cos \sqrt{2} \theta_s &=& \sin \left(\sqrt{2} \tilde{\phi}_s +
    \frac{h_s x}{u_s}\right) \\
  \sin \sqrt{2} \phi_s &=& - \cos \sqrt{2} \tilde{\theta}_s \\
  \cos \sqrt{2} \phi_s &=& \cos \left(\sqrt{2} \tilde{\phi}_s +
    \frac{h_s x}{u_s}\right).
\end{eqnarray}
Since for the rung current in the mean-field approximation we have:
\begin{eqnarray}
\nonumber
  J_\perp(x)&&=  \frac \Omega {\pi a} \left [ 2A_0^2 \sin
    (\pi \rho x -\sqrt{2}\theta_s -\gamma) -2 A_0 A_1 \sin
    (\sqrt{2}\theta_s +\sqrt{2}\phi_c + 2 \gamma - 2\pi \rho x )\cos (\sqrt{2}\phi_s ) \right. \\
  && -\left. 2 A_0 A_1 \sin  (\sqrt{2}\theta_s-2 \phi_c )\cos (\sqrt{2}\phi_s )\right]
 \end{eqnarray}
after the rotation we find:
\begin{eqnarray}
  J_\perp(x)&& = \frac{\Omega}{\pi a} \lbrack 2 A_0^2 \left (  \sin
    (\pi \rho x - \gamma) \sin (\sqrt{2}\tilde{\phi}_s + \frac{h_s
      x}{u_s}) -\cos (\pi \rho x -\gamma ) \sin (\sqrt{2}\tilde{\theta}_s\right )\nonumber \\
      && -2 A_0 A_1 \cos (2\gamma -2\pi \rho x) \sin \sqrt{2} \tilde{\theta}_s
      \cos \left(\sqrt{2} \tilde{\phi}_s + \frac{h_s x} {u_s} \right)
      \nonumber \\
     && + \sqrt{2} A_0 A_1 \sin (2\gamma -2 \pi \rho x) \partial_x
     \tilde{\phi}_s   \nonumber \\
     && -2 A_0 A_1\sin \sqrt{2} \tilde{\theta}_s
      \cos \left(\sqrt{2} \tilde{\phi}_s + \frac{h_s x} {u_s} \right) \rbrack,
\end{eqnarray}
we will have to take the average with respect to $\phi_s$ and
$\theta_s$ and with respect to $\gamma$. The latter averaging
partially takes into account the restablishment of the full $U(1)$
symmetry by fluctuations around the mean-field. Averaging over $\gamma$ gives
expressions that are translationally invariant. We find:}
\begin{eqnarray}
  \langle  J_\perp(j)  J_\perp(j')\rangle && \sim \Omega^2 A_0^2 A_1^2  \frac 1 {(j-j')^2}
  \cos \left(\frac{h_s (j-j')}{u_s} \right) \cos 2 \pi \rho (j-j')  +\nonumber \\
 && \Omega^2 A_0^2 A_1^2  \frac {\cos 2\pi \rho (j-j')}  {(j-j')^2} \nonumber \\
 && + \Omega^2 A_0^4\cos \pi \rho (j-j') \cos \left(\frac{h_s (j-j')}{u_s}
 \right)\frac{1}{|j-j'|}+ \Omega^2 A_0^4 \frac{\cos \pi \rho (j-j') }{|j-j'|}.
\end{eqnarray}

We therefore see that an incommensuration develops in the $k\simeq 0$ and $k\sim 2 \lambda=2\pi\rho$ component
of the rung-current and density-wave correlations.  In the Fourier transform peaks are located at $\pi \rho$ , $\pi \rho \pm h_s/u_s$ while the singularities
at  $2\pi \rho$ and  $2\pi \rho \pm h_s/u_s$  are discontinuities of slope.
Since $h_s \sim
\Omega^2$, the incommensuration increases with interchain hopping. One
can repeat the calculation also for the $S_z$ operator and its
correlation function $S_s(j-j') \sim \cos(\lambda j)\cos (\lambda
j')\frac{1}{|j-j'|}$, giving rise to a peak at $k \sim \pm \pi
\rho$. We note that since we have made very crude approximations to
treat the $\phi_c$ fluctuations, we cannot make accurate predictions
on the correct value of the exponents. \\
Regarding the calculation of the momentum distribution, since the boson annihilation operators do not correspond to primary fields of the SU(2)$_1$ WZNW model, we cannot
 derive their  expression in terms of $\tilde{\theta}_s$ and
 $\tilde{\phi_s}$ using the SU(2) rotation. However, since
 $e^{i\theta_s}$ has conformal dimensions (1/16, 1/16)  its expression
 in terms of the fields $\tilde{\phi}_s$ and $\tilde{\theta}_s$ can be
 expressed as a sum of operators of conformal dimensions (1/16,
 1/16). A general expression for the case $\lambda=\pi$ has been
 derived previously\cite{supplementary_prb}. The general form on $n_k$
 consisted of three peaks centered at $\pi \sigma \pm h_s/u_s$ and
$\pi \sigma$ or a single broad peak $\pi \sigma$, depending on the
value of $\Omega$. In the present case, a broad peak centered at
$2\lambda \sigma$ or a narrow peak at $2\lambda \sigma$ plus
satellites centered at $2\lambda
\sigma \pm h_s/u_s$  are expected. As we noted above, these results
can be derived rigorously provided we are at a commensurate filling and a charge gap
is formed. At generic filling, or when interactions in the charge
sector are insufficiently repulsive, the $U(1)$ symmetry of the
term~(\ref{eq:rung-pi-wzw}) is reestablished by quantum
fluctuations. In such case, the mean field treatment is only a
suggestion that incommensurate fluctuations will be present in a fully
gapless state. \\

\begin{figure}[h]
\begin{center}
\includegraphics[height=55.5mm]{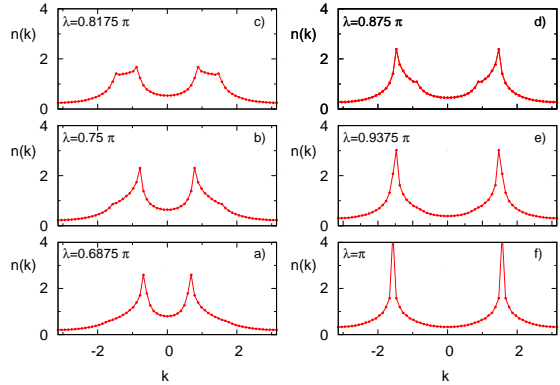}
\end{center}

\caption{Second incommensuration. DMRG simulation results at $L=64$ in PBC. Momentum distribution $n(k)$ at
$\rho=0.75$ for different values of the applied flux $\lambda$ as in the legend, spanning from below to well above the threshold $\lambda=0.75$ for the appearance of the second incommensuration. The interchain hopping is fixed at the value $\Omega/t=1.25$.}
\label{fig:nk_omega_n_0.75}
\end{figure}

In Fig.~\ref{fig:nk_omega_n_0.75} we follow the appearance of the second incommensuration in the
momentum distribution for the system at $\rho=0.75$, spanning from below the critical
$\lambda_c=\pi \rho=0.75 \pi$ up to $\lambda=\pi$, as from panels a) to f). At $\lambda=0.75\pi$ the appearance of the
secondary peaks are clearly detectable.

\begin{figure}[h]
\begin{center}
\includegraphics[height=55.5mm]{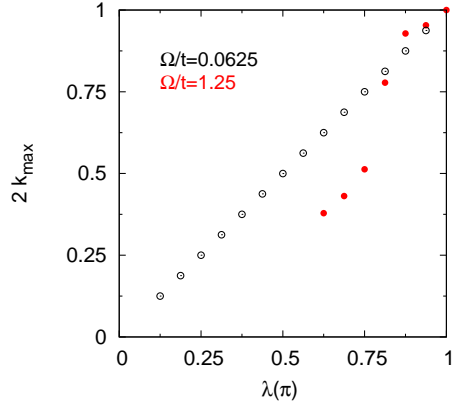}
\end{center}

\caption{Second incommensuration. DMRG simulation results at $L=64$ in PBC. Position  $k_{max}$ of the peaks of the momentum distribution $n(k)$
at $\rho=0.75$. Open black dots: $\Omega/t=0.0625$. Red solid dots: $\Omega/t=1.25$.}
\label{fig:pos_nk}
\end{figure}

The positions of these peaks move towards zero with increasing the
flux, and disappear completely at $\lambda=\pi$, where the system is back in the
standard Vortex phase. In the presence of the second incommensuration, the position of the peaks is no longer
proportional to the applied flux: this is apparent in Fig.~\ref{fig:pos_nk}, where the position $k_{max}$ of the peaks
in the momentum distribution is displayed as a function of $\lambda$ for the case with $\rho=0.75$ and $\Omega/t=1.25$
(red solid dots). For lower values of $\Omega/t$, the relation $q(\lambda)=\lambda$ is valid on
a large range, and the possible deviation at critical $\lambda$ is not appreciable, as it is seen in the
Fig.~\ref{fig:pos_nk}  (open black dots).

\begin{figure}[h]
\begin{center}
\includegraphics[height=55.5mm]{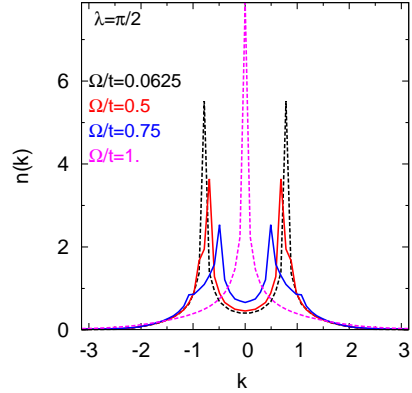}
\end{center}

\caption{Second incommensuration. DMRG simulation results at $L=64$ in PBC. Momentum distribution $n(k)$ at $\rho=0.5$ and fixed
applied flux $\lambda=\pi \rho $, for different values of $\Omega/t$ as in the legend. Dashed black line: $\Omega/t=0.0625$, where the system is in the standard vortex phase (first incommensuration). Dashed magenta line: $\Omega/t=1$, where the system is in the
Meissner phase. Red and blue solid lines: $\Omega/t=0.5$ and $\Omega/t=0.75$, respectively, where the occurrence of the second incommensuration is signaled by the appearance of the secondary peaks.}
\label{fig:nk_la16_n_0.5}
\end{figure}

We can also follow the evolution of the momentum distribution at the critical $\lambda_c=\pi \rho$ while varying the
interchain hopping $\Omega$. In Fig.~\ref{fig:nk_la16_n_0.5} we show $n(k)$ for the system at
$\rho=0.5$ and fixed applied flux $\lambda=\pi/2$ on increasing the interchain hopping. For the case with
$\Omega/t=0.0625$, represented by the dashed black line, the gap in total density is too small to be detected in the present numerical simulation at the $L=64$ system size. Yet, at $\Omega=0.5$ and $0.75$ the second incommensuration becomes
clearly visible with the predicted appearance of the secondary peaks.

\begin{figure}[h]
\begin{center}
\includegraphics[height=55.5mm]{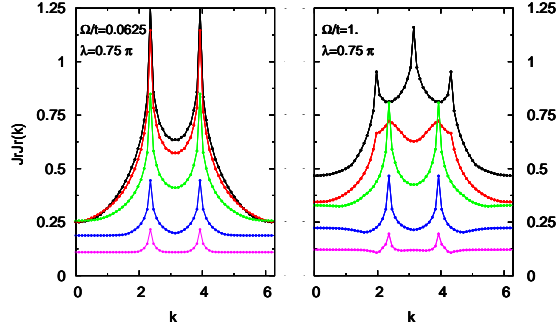}
\end{center}

\caption{First and second incommensuration. DMRG simulation results at $L=64$ in PBC. FT of the rung-current correlation function $C(k)$
at fixed applied flux $\lambda=0.75\pi$ for different fillings as in the legend: $\rho=1.0$, 0.75, 0.5, 0.25 and
0.125 are represented by black, red, green, blue and magenta solid lines, respectively. Left panel: case with $\Omega/t=0.0625$. Right panel: case with $\Omega/t=0.75$. Data at $\Omega/t=0.75$ and $\rho=1$ {has} been shifted to make more evident the
second incommensurations peaks.}
\label{fig:jrjrk}
\end{figure}

As mentioned above, the second incommensuration also shows up in the correlation
function for the rung-current. In Fig.~\ref{fig:jrjrk} we show the FT of this quantity at
$\lambda=0.75 \pi$ for different fillings.
The left panel of Fig.~\ref{fig:jrjrk} displays the data at a small value $\Omega/t=0.0625$: here, the system is
in the standard Vortex phase (first incommensuration) characterized by peaks located at $q(\lambda)=0.75 \pi$.
For the $\rho=0.75$ the small interchain hopping {leads to second
  incommensuration}  too small to be detected for the system size
of the present simulation. In all filling cases
the peaks are located at  $k=q(\lambda)=0.75 \pi$ and $k=2\pi-q(\lambda)=5/4\pi$.
The right panel of Fig.~\ref{fig:jrjrk} displays the data at $\Omega/t=0.75$, which is instead a sufficiently large value
so that the second incommensuration becomes sizable: indeed, $C(k)$ gets the expected second
{incommensuration} at the predicted filling $\rho=0.75$, while at smaller values of the filling no qualitative differences are seen with respect to the behavior shown in the left panel. At $\rho=1.0$ the second incommensuration appears at $\lambda=\pi$ and the peak at $k=\pi$, and it is still detectable for this applied flux\cite{our_2015}.

\begin{figure}[h]
\begin{center}
\includegraphics[height=55.5mm]{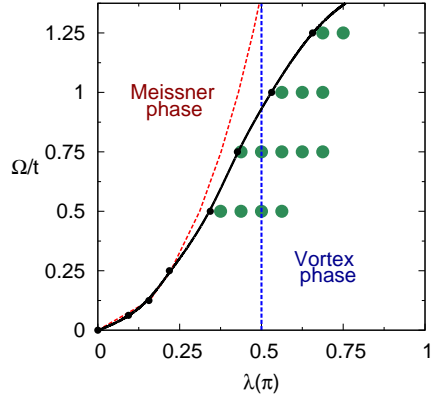}
\end{center}

\caption{DMRG simulation results at $L=64$ in PBC. Phase diagram for a hard-core bosonic system on ladder as a
function of flux per plaquette $\lambda$ and $\Omega/t$, at the filling value $\rho=0.5$. The occurrence of the two incommensurations is evidenced as follows. The black solid line represents the phase boundary between the Meissner and the
first incommensuration, a standard Vortex phase. The dark--green solid dots are the points where the
second incommensuration appears. The dashed blue line marks the critical $\lambda=0.5\pi$ at
which the second incommensuration is expected. For comparison, the phase boundary between the Meissner and
Vortex phase for a non-interacting system is represented as well, by the red-dashed line.
Notice the enhanced size of the Meissner region in the hard-core repulsive with respect to the non-interacting case.}
\label{fig:phd_n_0.5}
\end{figure}

In our previous study, we found\cite{our_2015} a large region of stability of the second incommensuration near
the critical value of $\lambda$. {In order to see how this region
evolves with the filling, we summarize in Fig.~\ref{fig:phd_n_0.5} the
phase diagram obtained from DMRG simulations in PBC for a system
size $L=64$ at  filling $\rho=0.5$, in which  the boundary in the transition from
Meissner to Vortex phase and the extension of the region near
$\lambda=\pi/2$  with the second incommensuration are visible. }

\begin{figure}[h]
\begin{center}
\includegraphics[height=55.5mm]{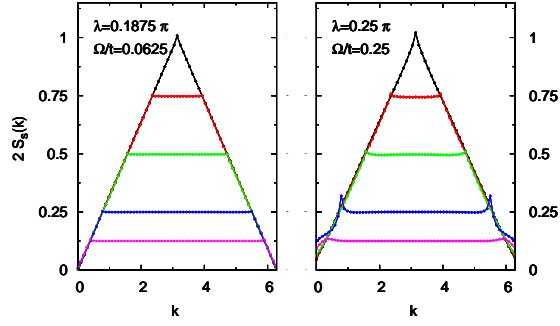}
\end{center}

\caption{DMRG simulation results at $L=64$ in PBC. Spin static structure factor $S_s(k)$ for different fillings as in the legend:
$\rho=1.0$, 0.75, 0.5, 0.25, and $0.125$ are represented by the black, red, green,
blue, and magenta solid lines, respectively. Left panel:
$\lambda=0.1875 \pi$ and the small value of interchain hopping $\Omega/t=0.0625$, where the system is always
in the standard Vortex phase at all fillings.
Right panel: $\lambda=0.25 \pi$ and $\Omega/t=0.25$. Notice that the system at $\rho=0.125$ is in the Meissner phase.}
\label{fig:sk}
\end{figure}

In Fig.~\ref{fig:sk} we show the behavior of $S_s(k)$ for different fillings in two different situations
in which only the first or also the second incommensuration appear.
In the left panel, the behavior in the standard vortex phase is displayed, after picking small values of $\lambda$
and $\Omega/t$. In the right panel, we show the behavior at $\lambda=0.25 \pi$: here, the appearance of the second incommensuration is expected at $\rho=0.25$, characterized by two peaks develop at $k=0.25\pi$ and $k=2\pi-0.25\pi$, and with  $S_s(k\to 0)$ getting a sizable finite value and a linear momentum dependence.
At the other
fillings, the spin correlation function
gets small finite values at $k=0$ and very low peaks at $k=\rho\pi$ and $k=2\pi-\rho\pi$, apart from the case $\rho=0.125$
already in the Meissner phase.

\begin{figure}[h]
\begin{center}
\includegraphics[height=55.5mm]{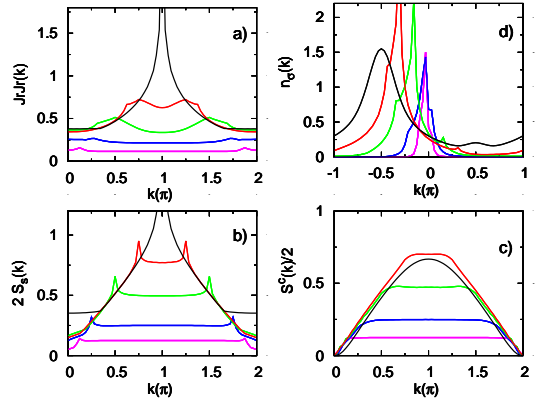}
\end{center}

\caption{DMRG simulation results at $L=64$ in PBC. Summary of the FT observables behavior at different fillings: $\rho=1.0$, 0.75, 0.5, 0.25, and $0.125$ are represented by the black,red,green,blue, and magenta solid lines, respectively.  The value of $\lambda$ is set to
$\lambda=\pi \rho$ and $\Omega/t$ is chosen in order to make the second incommensuration visible.
Panel a): rung-current correlation function $C(k)$. Panel b): spin correlation
function $S_s(k)$. Panel c): twice the charge correlation function $2S_c(k)$. Panel d): (spin-) chain-resolved momentum
distribution $n_\sigma(k)$, with $n_{-\sigma}(k)=n_{\sigma}(k)$. The different curve color represent different values of $\Omega/t$ as follows:
$\Omega/t=1.25$ (black), $\Omega/t=1$ (red), $\Omega/t=0.75$ (green), $\Omega/t=0.25$ (blue),
and $\Omega/t=0.0625$ (magenta).}
\label{fig:near}
\end{figure}

We conclude this section summarizing in Fig.~\ref{fig:near} the effects that the appearance of the second incommensuration produces in the different observables and quantities analyzed in the text. We see that there is almost no effect on
the charge static structure factor $S_c(k)$: as shown in
Fig.~\ref{fig:cic_n_0.75} or Fig.~\ref{fig:cic_n_0.5}, in fact no sharp slope discontinuity at $k=\rho\pi$ and $k=2\pi-\rho\pi$ emerges in the second incommensuration with respect to the first, that is the standard vortex case.
We remark the difference between the $\rho<1$ cases analyzed in the present
paper and the $\rho=1.0$ case\cite{our_2015}, which instead corresponds
to a Mott-insulator, {\it i.e.} quadratic behavior at low momenta.
The DMRG data show gapless leg-symmetric modes for  $\rho<1$ in
agreement with the
  Mermin-Wagner theorem\cite{mermin_wagner_theorem,hohenberg67_theorem}
   that implies no breaking of the $U(1)$ symmetry $\phi_c \to
   \phi_c+\gamma$, $\theta_s \to \theta_s +\gamma$.

\section{Conclusion}
{In conclusion, we have studied the commensurate-incommensurate
transition between the Meissner and the vortex state of a two-leg
bosonic ladder in an external flux, and the formation of a second
incommensuration in the vortex state when the flux is matching the
particle density.
The predictions of the bosonization treatment and  the results of DMRG
simulations on the commensurate-incommensurate transition from a commensurate Meissner
to a standard incommensurate Vortex phase, and the second incommensuration, have been discussed.
As expected from previous results at half-filling\cite{our_2015}, the
occurrence of a second incommensuration has been found  by the DMRG
simulations whenever the ratio between the flux and
the filling is equal to $\pi$. The developing
of the second incommensuration can be followed in the momentum distribution, that can be readily measured in experiments
\cite{atala2014}. A qualitative picture of the second
incommensuration, based on a phase averaging of mean-field
approximation of the bosonized theory has been presented.
Our DMRG results have been summarized in the interchain hopping-flux phase diagram Fig.~\ref{fig:phd_n_0.5}
at  quarter-filling, displaying the Meissner phase,
as well as  Vortex phase and second incommensuration.
The signatures of the second incommensuration on observables and correlation functions have been summed up
in Fig.~\ref{fig:near}.
Our predictions can be tested in current experiments, where observables that we have analyzed and discussed can be accessed.
A few questions remain open for future investigations. For example, one
could investigate whether a second incommensuration would also be
observed in multi-chain systems, such as a ladder with a few legs or a
two-dimensional array of bosonic chains. From the point of view of
bosonization, a more rigorous derivation of the second
incommensuration in the case of incommensurate filling
would be valuable.  }
\appendix

\section{Mapping to a spin ladder}
\label{sec:spinladder}

In the  hard core boson case, a representation\cite{fisher_xxz} in
terms of (pseudo) spins 1/2 can be introduced:
\begin{eqnarray}
  \label{eq:boson-to-spin}
  b_{j,\uparrow}^\dagger = S_{j,1}^+ \, b_{j,\downarrow}^\dagger = S_{j,2}^+,\\
 b_{j,\uparrow} = S_{j,1}^{-}\,   b_{j,\downarrow} = S_{j,2}^-,\\
 b^\dagger_{j,\uparrow} b_{j,\uparrow} = S_{j,1}^z+\frac 1 2\, b^\dagger_{j,\downarrow} b_{j,\downarrow} = S_{j,2}^z+\frac 1 2,
\end{eqnarray}
With such mapping, we can rewrite the
Hamiltonian~(\ref{eq:full-lattice-ham}) as a two-leg ladder
Hamiltonian:
\begin{eqnarray}
  \label{eq:dm-ladder}
  H&=&\sum_{j \atop r=1,2} J (S_{j,r}^x S_{j+1,r}^x +S_{j,r}^y
  S_{j+1,r}^y) + (-1)^r D (S_{j,r}^y S_{j+1,r}^x - S_{j,r}^x
  S_{j+1,r}^y) \\
&& + \sum_j  J_\perp (S_{j,1}^+ S_{j,2}^- +  S_{j,1}^+
S_{j,2}^-) + J_\perp^z S_{j,1}^z S_{j,2}^z \\
&& -\mu \sum_j   (S_{j,1}^z + S_{j,2}^z) + h  \sum_j   (S_{j,1}^z - S_{j,2}^z),
\end{eqnarray}
where $J=t \cos \lambda$, $D=t\sin \lambda$, $J_\perp=\frac \Omega 2$,
$J_\perp^z=U_{\uparrow \downarrow}$ and $h=\delta/2$. The term $D$ is
a uniform
Dzyaloshinskii-Moriya (DM)\cite{dzyalo_interaction,moryia_asym_int,kaplan_anisotropic,shekhtman_anisotropic_s}
interaction, with the DM vector parallel to $z$. For $\delta \ne 0$,
the two legs of the ladder are exposed to a different magnetic field.

\section{Fermionization approach}
\label{app:fermionization}

We have:
\begin{eqnarray}
  \label{eq:fermion}
  H_s=-i u_s \int dx (\psi^\dagger_R \partial_x
  \psi_R-\psi^\dagger_L \partial_x \psi_L)  - h  \int dx (\psi^\dagger_R  \psi_R+ \psi^\dagger_L
  \psi_L) -m \int dx (\psi^\dagger_R  \psi_L + \psi^\dagger_L \psi_R),
\end{eqnarray}
where $ m=\pi \Omega A_0^2 a$,$h=\frac{\lambda u_s}{2 a}$ and the
fermion annihilation operators $\psi_{R,L}^\dagger$ are destroying the
solitons. The detailed correspondence between the fermionic and bosonic
expression of the lattice operators is derived below.

The fermionized Hamiltonian (\ref{eq:fermion}) is obtained by the following correspondence with the boson operators:
\begin{eqnarray}\label{eq:fermionization}
  \psi^\dagger_R  \psi_R+ \psi^\dagger_L \psi_L &=& -\frac{\partial_x \theta_s}{\pi \sqrt{2}} \\
 \psi^\dagger_R  \psi_R- \psi^\dagger_L \psi_L&=&\frac{\sqrt{2} \partial_x \phi_s}{\pi} \\
 \psi^\dagger_R  \psi_L + \psi^\dagger_L \psi_R &=& \frac{\cos \sqrt{2} \theta_s}{\pi \alpha} \\
 -i(\psi^\dagger_R  \psi_L - \psi^\dagger_L \psi_R) &=& \frac{\sin \sqrt{2} \theta_s}{\pi \alpha}
\end{eqnarray}

Within fermionization approach the single-particle correlation function $ \langle \psi^\dagger_R \psi_L\rangle$ can be evaluated by the single-particle Green's function of the operators that diagonalize the Hamiltonian (\ref{eq:fourier})

\begin{eqnarray}
  \label{eq:RL-expectation}
  \langle \psi^\dagger_R \psi_L\rangle &=& \int \frac{dk}{2\pi} \sin \varphi_k \cos \varphi_k \langle c^\dagger_{k,+} c_{k,+} -  c^\dagger_{k,-} c_{k,-}\rangle \\
&=& \int_{\sqrt{h^2-\Delta^2}/u}^{\Lambda} \frac{dk}{2\pi} \frac{m}{\sqrt{uk}^2 + m^2} \\
&=& \frac{m}{4\pi v} \ln\left(\frac{2v\Lambda}{h+\sqrt{h^2-m^2}}\right).
\end{eqnarray}
since $\langle \psi^\dagger_R \psi_L\rangle$ is real, there is no average current between the legs of the ladder.
However, it is possible to find fluctuations of the current as shown in Eq.\ref{eq:wick_res}.
In the commensurate phase the correlators can be evaluated using (\ref{eq:fourier}) and one obtains

 \begin{eqnarray}
  \langle \psi^\dagger_{R}(x) \psi_L(x')\rangle = -\int_{-\infty}^\infty  \frac{dk}{4\pi} \frac{m}{\sqrt{(uk)^2+m^2}} e^{ik(x'-x)}, \\
   \langle \psi^\dagger_{L}(x) \psi_R(x')\rangle = -\int_{-\infty}^\infty  \frac{dk}{4\pi} \frac{m}{\sqrt{(uk)^2+m^2}} e^{ik(x'-x)}, \\
   \langle \psi^\dagger_{R}(x) \psi_R(x')\rangle = -\int_{-\infty}^\infty  \frac{dk}{4\pi} \frac{uk}{\sqrt{(uk)^2+m^2}} e^{ik(x'-x)}, \\
   \langle \psi^\dagger_{L}(x) \psi_L(x')\rangle = \int_{-\infty}^\infty  \frac{dk}{4\pi} \frac{uk}{\sqrt{(uk)^2+m^2}} e^{ik(x'-x)},
\end{eqnarray}
so that:
\begin{eqnarray}
  \langle \psi^\dagger_{R}(x) \psi_L(x')\rangle =  \langle \psi^\dagger_{L}(x) \psi_R(x')\rangle &=& -\frac{m}{2\pi u} K_0\left(\frac{m |x-x'|}{u}\right), \\
 \langle \psi^\dagger_{R}(x) \psi_R(x')\rangle = - \langle \psi^\dagger_{L}(x) \psi_L(x')\rangle &=& i \mathrm{sign}(x-x')  \frac{m}{2\pi u} K_1\left(\frac{m |x-x'|}{u}\right),
\end{eqnarray}
where $K_0$ and $K_1$ are modified Bessel functions. This leads to the result (\ref{eq:corr_res}) and one expects an exponential decay of the rung current correlation with correlation length $u/(2m)$. \\

In the incommensurate phase, the fermion correlation functions are
expressible instead in terms of incomplete Bessel functions\cite{agrest1971}:
\begin{equation}
  \epsilon_\nu(w,z)=\frac 1 {i\pi} \int_0^w e^{z \sinh t -\nu t} dt
\end{equation}

Indeed, we have (for $T=0$):
\begin{eqnarray}
  \label{eq:correlators-incom}
  \langle \psi_R^\dagger(x) \psi_L(x')\rangle &=& \int_{-k_F}^{k_F} \frac{dk}{4\pi} \frac{m}{\sqrt{(uk)^2+m^2}} e^{ik (x'-x)} - \int_{-\infty}^{\infty} \frac{dk}{4\pi} \frac{m}{\sqrt{(uk)^2+m^2}} e^{ik (x'-x)},  \\
  \langle \psi_R^\dagger(x) \psi_R(x')\rangle &=&  \int_{-k_F}^{k_F} \frac{dk}{4\pi} \left(1+ \frac{uk}{\sqrt{(uk)^2+m^2}}\right) e^{ik (x'-x)} - \int_{-\infty}^{\infty} \frac{dk}{4\pi} \frac{uk}{\sqrt{(uk)^2+m,^2}} e^{ik (x'-x)},  \\
 \langle \psi_L^\dagger(x) \psi_L(x')\rangle &=&  \int_{-k_F}^{k_F} \frac{dk}{4\pi} \left(1- \frac{uk}{\sqrt{(uk)^2+m^2}}\right) e^{ik (x'-x)} + \int_{-\infty}^{\infty} \frac{dk}{4\pi} \frac{uk}{\sqrt{(uk)^2+m^2}} e^{ik (x'-x)},
\end{eqnarray}
where $\sqrt{(uk_F)^2+m^2}=h$ and we have noted that $\langle \psi_R^\dagger(x) \psi_L(x')\rangle=\langle \psi_L^\dagger(x) \psi_R(x')\rangle$.
In Eq.(\ref{eq:correlators-incom}), we have two contributions, one coming from the partially filled upper band, and the other from the filled lower band which was the only contribution in the commensurate case. We see that when $h \to +\infty$, $k_F\to \infty$ and   $\langle \psi_R^\dagger(x) \psi_L(x')\rangle \to 0$ uniformly. We have:
\begin{eqnarray}
   \int_{-k_F}^{k_F} \frac{dk}{4\pi} \frac{m}{\sqrt{(uk)^2+m^2}} e^{ik (x'-x)}&=&\frac m u \int_{-\theta_F}^{\theta_F} \frac{d\theta}{4\pi} e^{i \frac{m (x'-x)}{u} \sinh \theta}\nonumber\\
&=& \frac {im}{4u} \left[\epsilon_0\left(\theta_F,i\frac{m(x'-x)} u\right) - \epsilon_0\left(-\theta_F,i\frac{m(x'-x)} u\right)\right] \\
 \int_{-k_F}^{k_F} \frac{dk}{4\pi} \frac{uk}{\sqrt{(uk)^2+m^2}} e^{ik (x'-x)}&=&\frac m u \int_{-\theta_F}^{\theta_F} \frac{d\theta}{4\pi} e^{i \frac{m (x'-x)}{u} \sinh \theta}\sinh \theta \nonumber\\
&=& \frac {im}{8u} \left[\epsilon_1\left(\theta_F,i\frac{m(x'-x)} u\right)- \epsilon_{-1}\left(\theta_F,i\frac{m(x'-x)} u\right) - \epsilon_1\left(-\theta_F,i\frac{m(x'-x)} u\right)\right.\nonumber \\ && \left. +\epsilon_{-1}\left(-\theta_F,i\frac{m(x'-x)} u\right)\right]
\end{eqnarray}
and thus:
\begin{eqnarray}
  \label{eq:incom-bessel-corr}
  \langle \psi_R^\dagger(x) \psi_L(x')\rangle &=&  \frac {im}{4u} \left[\epsilon_0\left(\theta_F,i\frac{m(x'-x)} u\right) - \epsilon_0\left(-\theta_F,i\frac{m(x'-x)} u\right)\right] -\frac{m}{2\pi u} K_0\left(\frac{m |x-x'|}{u}\right) \\
 \langle \psi_R^\dagger(x) \psi_R(x')\rangle &=& \frac{\sin k_F(x'-x)}{2\pi (x'-x)} +  \frac {im}{8u} \left[\epsilon_1\left(\theta_F,i\frac{m(x'-x)} u\right)- \epsilon_{-1}\left(\theta_F,i\frac{m(x'-x)} u\right) - \epsilon_1\left(-\theta_F,i\frac{m(x'-x)} u\right)\right. \nonumber\\ && \left.+\epsilon_{-1}\left(-\theta_F,i\frac{m(x'-x)} u\right)\right] + i \mathrm{sign}(x-x')  \frac{m}{2\pi u} K_1\left(\frac{m |x-x'|}{u}\right)\nonumber \\
\langle \psi_L^\dagger(x) \psi_L(x')\rangle &=& \frac{\sin k_F(x'-x)}{2\pi (x'-x)} -  \frac {im}{8u} \left[\epsilon_1\left(\theta_F,i\frac{m(x'-x)} u\right)- \epsilon_{-1}\left(\theta_F,i\frac{m(x'-x)} u\right) - \epsilon_1\left(-\theta_F,i\frac{m(x'-x)} u\right)\right. \nonumber\\ && \left.+\epsilon_{-1}\left(-\theta_F,i\frac{m(x'-x)} u\right)\right] - i \mathrm{sign}(x-x')  \frac{m}{2\pi u} K_1\left(\frac{m |x-x'|}{u}\right)
\end{eqnarray}

For large distances, $|x-x'| \gg u/m$, we can neglect the contribution from the lower band. The contribution from the upper band can be obtained from the asymptotic expansions given in Ref.~\onlinecite{agrest1971} on p. 146, while the simpler derivation can be obtained from physical arguments and is presented in the main text. Indeed, in the case $ u k_F \ll m$,
we can make the approximations:
\begin{eqnarray}
  \frac{m}{\sqrt{(uk)^2 + m^2}} \simeq 1 \\
   \frac{uk}{\sqrt{(uk)^2 + m^2}} \simeq \frac{uk}{m}
\end{eqnarray}
giving:
\begin{eqnarray}
  \langle \psi_R^\dagger(x) \psi_L(x')\rangle \simeq \frac{\sin k_F(x'-x)}{2\pi (x'-x)} + O(x-x')^{-3} \\
  \langle \psi_R^\dagger(x) \psi_R(x') \rangle = \frac{\sin k_F(x'-x)}{2\pi (x'-x)} -\frac{iuk_F}{m} \left[\frac{\cos k_F(x'-x)}{2\pi (x'-x)} +\frac{\sin k_F(x'-x)}{2\pi k_F (x'-x)^2}\right] + O(x-x')^{-3} \\
 \langle \psi_L^\dagger(x) \psi_L(x') \rangle = \frac{\sin k_F(x'-x)}{2\pi (x'-x)} +\frac{iuk_F}{m} \left[\frac{\cos k_F(x'-x)}{2\pi (x'-x)} +\frac{\sin k_F(x'-x)}{2\pi k_F (x'-x)^2}\right] + O(x-x')^{-3}
\end{eqnarray}

Second, in the case of $u k_F \gg m$, we can linearize the dispersion in the upper band around the points $\pm k_F$.
We can then make the approximations:
\begin{eqnarray}
  \frac{u (k\pm k_F)}{\sqrt{u^2(k\pm k_F)^2+m^2}} \simeq \pm 1
\end{eqnarray}
This time, we find:
\begin{eqnarray}\label{eq:incom-large-kf}
  \langle \psi^\dagger_R(x) \psi_R(x')\rangle &=& e^{i k_F (x'-x)}\int_{-\infty}^0 \frac{dk}{2\pi} e^{k(\alpha+i(x'-x))} = \frac{e^{i k_F(x'-x)}}{2 \pi [\alpha+i(x'-x)]} \\
    \langle \psi^\dagger_L(x) \psi_L(x')\rangle &=& e^{-i k_F (x'-x)}\int^{+\infty}_0 \frac{dk}{2\pi} e^{k(-\alpha+i(x'-x))} = \frac{e^{-i k_F(x'-x)}}{2 \pi [\alpha-i(x'-x)]} \\
    \langle \psi^\dagger_R(x) \psi_L(x')\rangle &=& \frac{m}{2\pi u k_F} \frac{\sin k_F(x'-x)}{(x'-x)}
\end{eqnarray}
We see that the correlator $ \langle \psi^\dagger_R(x) \psi_L(x')\rangle$ is smaller by a factor $m/(u k_F) \sim m/h \ll 1$ in that limit. If we had instead written a bosonized
Hamiltonian, we would have found that $\langle \psi^\dagger_R(x) \psi_L(x')\rangle =0$.
With Eqs.(\ref{eq:incom-large-kf}), we obtain the expression for the rung-current correlator (\ref{eq:corr_res_2}).

In the fermionic representation~(\ref{eq:fermion}), the Hamiltonian is
readily diagonalized in the form
\begin{equation}
  \label{eq:fermi-diag}
  H=\sum_{k,r=\pm}  (r \sqrt{(u_s k)^2+m^2} - h) c^\dagger_{k,r} c_{k,r},
\end{equation}
by writing:
\begin{equation}
  \label{eq:fourier}
  \left(\begin{array}{c}
    \psi_R(x) \\ \psi_L(x)
  \end{array}\right) = \frac 1 {\sqrt{L}} \sum_k e^{ikx} \left(\begin{array}{cc}
\cos \varphi_k & -\sin \varphi_k \\
\sin \varphi_k & \cos \varphi_k
  \end{array} \right) \left(\begin{array}{c} c_{k+} \\ c_{k-} \end{array}\right),
\end{equation}
with: $e^{2i \varphi_k} =\frac{uk+im}{\sqrt{(uk)^2+m^2}}$. The
commensurate phase\cite{japaridze_cic_transition} is obtained for $|h|<|m|$ and the incommensurate
phase for $|h|>|m|$.

We can express the currents as:
\begin{eqnarray}\label{eq:fermi-jpar}
  J_\parallel(\lambda)&=&\frac{u_s}{2}\left[\frac{\lambda}{2\pi a} -
    (\psi^\dagger_R \psi_R + \psi^\dagger_L \psi_L)\right],  \\
\label{eq:fermi-jperp}
 J_\perp(\lambda) &=&  i \sqrt{2} m  (\psi^\dagger_R \psi_L - \psi^\dagger_L \psi_R),
\end{eqnarray}
and the $q\sim 0$ component of $n_{j\uparrow} -n_{j\downarrow}$ as:
\begin{eqnarray}
  (n_{j\uparrow} -n_{j\downarrow})_{q\sim 0} \sim  \psi^\dagger_R
  \psi_R - \psi^\dagger_L \psi_L.
\end{eqnarray}
Using (\ref{eq:fermi-jpar}), one has\cite{orignac01_meissner} $\langle J_\parallel \rangle=\frac{u_s \lambda}{4\pi a}$  in the commensurate phase, and
\begin{eqnarray}
  \langle J_\parallel \rangle  = \frac{u_s}{4\pi a}\left[\lambda - \sqrt{\lambda^2
    -\lambda_c^2}\right]
\end{eqnarray}
in the incommensurate phase. The finite-size scaling of the leg current has been derived
in\cite{didio2015a}.
As $ \langle \psi^\dagger_R \psi_L\rangle$ is real, the average rung current vanishes.

However, rung-current fluctuations are non-vanishing.
Indeed, with the help of Wick's theorem we obtain:
\begin{eqnarray}
\label{eq:wick_res}
  \langle J_\perp(x) J_\perp(x')\rangle &\propto& \left[\langle
    \psi^\dagger_{R} (x) \psi_L(x')\rangle  \langle\psi^\dagger_{R} (x')
    \psi_L(x)\rangle +   \langle \psi^\dagger_{L} (x)
    \psi_R(x')\rangle  \langle\psi^\dagger_{L} (x') \psi_R(x)\rangle
  \right. \\ && \left. -  \langle \psi^\dagger_{R} (x)
    \psi_R(x')\rangle  \langle\psi^\dagger_{L} (x') \psi_L(x)\rangle -
    \langle \psi^\dagger_{L} (x) \psi_L(x')\rangle
    \langle\psi^\dagger_{R} (x') \psi_R(x)\rangle\right].
\end{eqnarray}
In the commensurate phase, the correlators in (\ref{eq:wick_res}) can be evaluated using (\ref{eq:fourier}). One obtains:
\begin{eqnarray}
\label{eq:corr_res}
  \langle J_\perp(x) J_\perp(x')\rangle \propto \left(\frac{m}{2\pi u_s}\right)^2 \left[ K_0\left(\frac{m |x-x'|}{u_s}\right)^2 +  K_1\left(\frac{m |x-x'|}{u_s}\right)^2\right],
\end{eqnarray}
where $K_0$ and $K_1$ are the modified Bessel functions. The
exponential decay is thus recovered for $|x-x'|\gg u/m$. Taking the Fourier
transform, we find
\begin{eqnarray}
  C(0) -C(k)  \propto  E\left(-\frac{(uk)^2}{(2m)^2}\right) -
  K\left(-\frac{(uk)^2}{(2m)^2}\right),
\end{eqnarray}
where $E$ and $K$ are complete elliptic
integrals\cite{abramowitz_math_functions}. Using the
fermion representation, we can also show that:
\begin{eqnarray}
   S_s(0)-S_s(k) \sim \pi - E\left(-\frac{(uk)^2}{(2m)^2}\right).
\end{eqnarray}
In the incommensurate phase, the fermion correlation functions are
expressible instead in terms of incomplete Bessel
functions\cite{agrest1971}. The detailed expressions are reported in
the Appendix \ref{app:fermionization}).
For large distances, $|x-x'| \gg u/m$, we can neglect the contribution
from the lower band. The contribution from the upper band can be
obtained from the asymptotic expansions given in
Ref.~\onlinecite{agrest1971}. In the limit $u k_F \gg m$, simple
physical arguments give:
\begin{eqnarray}
\label{eq:corr_res_2}
  \langle J_\perp(x) J_\perp(x')\rangle \sim \frac{\cos
    2k_F(x-x')}{4\pi^2 (x-x')^2} +\ldots ,
\end{eqnarray}
so the Fermi wavevector $k_F=\sqrt{h^2-m^2}/u_s=q(\lambda)/2$.
Taking the Fourier transform (\ref{eq:corr_res_2}), we deduce that
$C(k)$ has slope discontinuities at $k=\pm 2k_F$.  By contrast, in
that limit, we find that
$\langle
(n_{j\uparrow}-n_{j\downarrow})(n_{j'\uparrow}-n_{j'\downarrow})\rangle
\sim (j-j')^2$ as expected from the bosonization arguments.


 \end{document}